\documentclass[lettersize,journal]{IEEEtran}
\usepackage{amsmath,amsfonts}
\usepackage{algorithmic}
\usepackage{algorithm}
\usepackage{array}
\usepackage[caption=false,font=footnotesize,labelfont=rm,textfont=rm]{subfig}
\usepackage{textcomp}
\usepackage{stfloats}
\usepackage{url}
\usepackage{verbatim} 
\usepackage{graphicx}  		%插入图片的宏包
\usepackage{float}		  	%设置图片浮动位置的宏包
\usepackage{subfig} 
\usepackage{subfloat}
\usepackage{stfloats}
\usepackage{cite}
\usepackage{bm}
\usepackage{amssymb}
\usepackage{makecell}
\usepackage{color}
\usepackage{soul}			%高亮文本
\usepackage[backref]{hyperref}
\usepackage{booktabs}
\usepackage{multirow}
\usepackage{lscape}

\begin{document}
	
	\title{Graph Embedding Dynamic Feature-based Supervised Contrastive Learning of Transient Stability for Changing Power Grid Topologies}
	
	\author{Zijian~Lv, Xin~Chen*, Zijian~Feng
		\thanks{This work was supported in part by the National Natural Science Foundation of China (Grant No.21773182 (B030103)) and the HPC Platform, Xi'an Jiaotong University.}
		\thanks{ Zijian~Lv (e-mail: jtlzj271828@stu.xjtu.edu.cn) ,  Xin Chen (Corresponding author, e-mail: xin.chen.nj@xjtu.edu.cn), and Zijian~Feng (e-mail: KenFeng@stu.xjtu.edu.cn)  are with the School of Electrical Engineering and the Center of Nanomaterials for Renewable Energy, State Key Laboratory of Electrical Insulation and Power Equipment, School of Electrical Engineering, Xi'an Jiaotong University, Xi'an, Shaanxi, China .}
	}

	\maketitle
	
	\begin{abstract}
		Accurate online transient stability prediction is critical for ensuring power system stability when facing disturbances. While traditional transient stablity analysis replies on the time domain simulations can not be quickly adapted to the power grid toplogy change. In order to vectorize high-dimensional power grid topological structure information into low-dimensional node-based graph embedding streaming data, graph embedding dynamic feature (GEDF) has been proposed. The transient stability GEDF-based supervised contrastive learning (GEDF-SCL) model uses supervised contrastive learning to predict transient stability with GEDFs, considering power grid topology information. To evaluate the performance of the proposed GEDF-SCL model, power grids of varying topologies were generated based on the IEEE 39-bus system model. Transient operational data was obtained by simulating N-1 and N-$\bm{m}$-1 contingencies on these generated power system topologies. Test result demonstrated that the GEDF-SCL model can achieve high accuracy in transient stability prediction and adapt well to changing power grid topologies.
		
	\end{abstract}
	
	\begin{IEEEkeywords}
		
		Transient stability assessment, topological embedding, supervised contrastive learning, topological change.
		
	\end{IEEEkeywords}
	
	\section{Introduction}
	
	\IEEEPARstart{A} reliable power system requires the steady synchronous operation of synchronous generators to meet load demands\cite{Pavella2000}. Therefore, power systems must withstand shocks, including common disturbances such as sudden generator failures, transmission line faults, and ground faults. Modern power systems have become increasingly complex due to large-scale grid-connected renewable energy and the widespread use of power electronic devices, making systems more susceptible to various potential disturbances\cite{osti_1220662,9027270}.Transient stability refers to the ability of synchronous generators to maintain synchronism after disturbances occur\cite{KARAMI2011983}. Transient stability assessment (TSA) in power systems is an essential means for rapidly assessing and maintaining stable operations. Meanwhile, transient stability prediction (TSP) is another critical concept focusing on rapid post-fault TSA to anticipate potential transient instabilities\cite{8920121}. The analysis of power systems poses undoubted challenges, considering the complexity arising from higher-order nonlinear dynamics and the intricate topology of power grids. At the same time, transient processes in power systems typically unfold over an extremely short period (less than a second)\cite{Pavella2000}, posing critical challenges for real-time assessment and control.
	
	Existing model-driven methods for TSA include time domain simulation (TDS)\cite{TDS}, transient energy function (TEF) techniques\cite{TEF}, and approaches based on Lyapunov exponents (LEs)\cite{LEs}. TDS, which involves the solution of differential-algebraic equations modeling both the generator and the power system topology, is typically known for its accuracy and reliability\cite{8382312}. However, the time-consuming nature of TDS calculations makes them unsuitable for real-time prediction of transient stability\cite{1994Power}. TEF and Extended Equal Area methods are useful, but constructing their energy functions is often difficult, making them less suited for some TSA applications\cite{COLVARA2009139}. Alternatively, the LEs approach has proven effective for determining transient stability by calculating the maximum Lyapunov exponent (MLE) of the system based on phasor measurement unit (PMU) data\cite{LEs}. However, due to limitations in the observation time window length, the MLE calculation using the LEs method takes several seconds\cite{6939746}.
	
	The establishment of wide area measurement systems (WAMS) utilizing PMUs has enabled synchronized measurements, leading to new concepts in control. This has led to numerous data-driven TSA approaches leveraging pattern recognition and machine learning techniques\cite{7932885, ANN, ANN_2, SVM, SVM_2, SVM_3, DT, RF}. In general, machine learning algorithms for TSA do not rely on prior knowledge or physical models of the power system, depending entirely on data for training. While this grants them the ability to effectively analyze large datasets and extract salient features, it also imposes greater demands on the quantity and quality of training data.
	
	Among various deep learning networks currently in use, convolutional neural networks (CNNs) have demonstrated significant success in image recognition. CNNs have also been explored and applied to transient stability analysis of power systems. For example, Hou \textit{et al.}\cite{8403460} proposed a voltage phase complex plane imaging method. Using images as input, a model for fast assessment of power system transient stability based on CNN was proposed. In addition, Zhu \textit{et al.}\cite{8920121} proposed preprocessing data with Fourier transforms to extract critical features, stitching them into 2D images to train a CNN.
	
	Graph convolutional networks (GCNs) seamlessly integrate topological structure and convolutional computation, making them well-suited for adapting to the data structure of power systems\cite{Sun2022FastTS}. However, GCNs face two significant challenges. The first is the sparsity of the power system's adjacency matrix, which creates major inefficiencies in memory usage and computations as the system scales. To address this, Huang \textit{et al.}\cite{9081965} proposed treating batches of graphs as subgraphs of one or more larger graphs, utilizing parallel computing to reduce computational complexity. The second relates to the inability to perform inductive tasks, {\it{i.e.}} dealing with dynamic graph problems. The effectiveness of the trained GCN model can be compromised if the grid topology changes. The general strategy is to stop the running GCN model and retrain it offline. To reduce the complexity of the transfer training process, Li and Wu\cite{Pavella2000} proposed using maximum mean discrepancy (MMD) to identify the most challenging samples for classification by calculating the differences between the two datasets. Zhang \textit{et al.}\cite{9942331} utilized the local maximum mean discrepancy (LMMD) method, which focuses on local differences compared to MMD. LMMD is more stable and reliable computationally. Similarly, it reduced transfer training costs and achieved improved performance.
	
	In summary, the main contributions of this paper are as follows:
	\begin{enumerate}
		\item{A physics-informed node2vec-based graph embedding method is constructed to encode the transient PMU data with the power grid topology information to generate the node-based graph embedding dynamic features (GEDFs).}
		\item{Based on GEDFs and a data augmentation technique, the transient stability GEDF-based supervised contrastive learning (GEDF-SCL) model is proposed based on the Supervised Contrastive Learning, which is tasked with TSP of the power systems with changing power grid topologies.}
		\item{GEDF-SCL has the 94.39\% TSP accuracy for the N-1 contingency based on the dataset of 39 bus power systems with changing topologies, The GEDF-SCL model is further examined for the N-$m$-1 contingency by fine-tuning the trained model from the N-1 contingency. TSP accuracy can reach 97.51\%, 96.59\% and 93.75\% for the N-1-1, N-2-1, and N-3-1 contingencies.}
	\end{enumerate}
	
	The paper is organized as follows: In Section~\ref{GEDFs}, the GEDFs construction is introduced. In Section~\ref{GEDFL} describes the structure and training methods of GEDF-SCL, including supervised comparison learning and data augmentation methods. In Section~\ref{N-1_test}, the stability prediction performance of the GEDF-SCL model is tested with N-1 contingency scenarios. In Section~\ref{N-$m$-1_test}, the model's transferability to the N-$m$-1 contingency is tested. Conclusions are drawn in Section~\ref{conclusion}.
	
	\section{Physics-informed Graph Embedding Dynamic Features}
	\label{GEDFs}
	
	The physics-informed graph embedding dynamic features (GEDFs) method is proposed to accurately encode the power grid topology information and transient power flow in power systems, effectively mapping high-dimensional data from complex structures into a low-dimensional feature space.
	
	The power system is modeled as a weighted and undirected graph $G(\mathcal{V};\mathcal{E};\bm{\mathcal{A}})$ with $n$ nodes (or buses) $\mathcal{V}=\{1,\cdots,n\}$, edges (or branches) $\mathcal{E}$, and edge weights $\bm{\mathcal{A}}\in \mathbb{R}^{n\times n}$. At time $t$, the bus $i$ is associated with a phasor voltage $U_t^i=V_t^ie^{\textbf{j}\theta_t^i}\in \mathbb{C}$, where $V_t^i\geq 0$ is the voltage magnitude and $0\leq \theta_t^i \leq 2\pi$ is the voltage angle, and $\bm{\theta}_t=(\theta_t^1, \theta_t^2,\cdots,\theta_t^n)^T$ denotes the voltage angle of all nodes. Transmission lines are represented using the standard lumped parameter $\Pi$-model, which accommodates parallel inductive/capacitive elements, tap-changing transformers, and line-charging capacitors. The impedance between node $i$ and node $j$ is defined as $1/y_{ij}=r_{ij}+jx_{ij}$, where $r_{ij}$ is the transmission line resistance and $x_{ij}$ is the reactance. Usually, in power systems, especially high-voltage systems, the resistance of the  transmission lines is neglected due to its relatively small size\cite{2012Power}, and it is often neglected in transient stability studies. If line charging and parallel capacitance are ignored, then $y_{ij}=1/x_{ij}$.
	
	The adjacency matrix $\bm{\mathcal{A}}$ represents the weights and topological structure of the power system, where the elements are defined as $\mathcal{A}_{ij}=-V^i V^j y_{ij}$ and $\mathcal{A}_{ii}=-\sum_{j=0,j\neq i}^{n}\mathcal{A}_{ij}$. Matrix $\bm{\mathcal{A}}$ can be written as $\bm{\mathcal{A}} = \bm{B} \text{diag}(\{a_{ij}\}_{{i,j}\in \mathcal{E}}) \bm{B}^T$\cite{Godsil2001}. The term $\text{diag}(\{a_{ij}\}_{{i,j}\in \mathcal{E}})$ constitutes a diagonal matrix with $a_{ij}=-V_i V_j y_{ij}$ as diagonal elements, while $\bm{B}$ is the (oriented) incidence matrix. For the $l$-th edge, where $l\in \{1,\cdots,\vert \mathcal{E}\vert\}$, and the edge direction can be arbitrary, the matrix $\bm{B}\in \mathbb{R}^{n\times \vert \mathcal{E}\vert}$ can be defined element-wise as: $B_{kl}=1$ if node $k$ is the sink node of edge $l$, $B_{kl}=-1$ if node $k$ is the source node of edge $l$, and all other elements are 0. The transient active power in power systems at time $t$ can be calculated by\cite{8347206,doi:10.1073/pnas.1212134110}
	\begin{equation}
		\label{P}
		\mathbf{p}_t = \bm{B}\text{diag}\left(\{a_{ij}\}_{\{i,j\}\in\mathcal{E}}\right)\sin(\bm{B}^T\boldsymbol{\theta}_t).
	\end{equation}
	By replacing each term $\sin(B^T\boldsymbol{\theta}_t)$ with the auxiliary vector $\bm{\psi}_{t} \in \mathbb{R}^{\vert\mathcal{E}\vert}$ , Eq.~\eqref{P} is can be rewritten as
	\begin{subequations}\label{eq:5}
		\begin{align}
			&\mathbf{p}_t = \bm{B}\text{diag}(\{a_{ij}\}_{\{i,j\}\in\mathcal{E}})\bm{\psi}_t,             \\
			&\bm{\psi}_t = \sin(\bm{B}^T\boldsymbol{\theta}_t).
		\end{align}
	\end{subequations}
	Eq.~\eqref{eq:5} essentially is the auxiliary-fixed point equation\cite{doi:10.1073/pnas.1212134110}. According to the definition of the adjacency matrix $\bm{\mathcal{A}}$, there exists a zero eigenvalue, indicating that the adjacency matrix is singular. Using the Moore-Penrose pseudo-inverse matrix $\bm{\mathcal{A}}^\dagger$\cite{doi:10.1073/pnas.1212134110}, the vector $\bm{\psi}_t$ of Eq.~\eqref{eq:5} can be calculated by
	\begin{equation}
		\label{eq6}
		\bm{\psi}_t = \bm{B}^T \bm{\mathcal{A}}^\dagger \mathbf{p}_t.
	\end{equation}
	
	The GEDF vector is defined as
	\begin{equation}
		\label{feature_map}
		\boldsymbol{\Delta}_t \overset{\underset{\mathrm{def}}{}}{=} \bm{\mathcal{A}}^\dagger \mathbf{p}_t.
	\end{equation}
	If the sampling period is $dT$, the total sampling time is $T$, and the total number of sampling points is $N=T/dT$, the GEDFs $\bm{\mathcal{F}}_N$ of size $(n\times N)$ can be obtained as
	\begin{equation}
		\label{GEDF_def}
		\bm{\mathcal{F}}_N = \begin{pmatrix}
			\boldsymbol{\Delta}_{0}, & \boldsymbol{\Delta}_{1},
			& \cdots, & \boldsymbol{\Delta}_{N}
		\end{pmatrix}.
	\end{equation}
	
	When the power system reaches a steady state, all phase angles are in the region of a synchronization manifold of the coupled oscillator model with an infinite number of possible solutions\cite{doi:10.1073/pnas.1212134110}. 
	Whereas the phase angle difference, {\it{i.e.}} $\mathbf{B}^T\boldsymbol{\theta}^*$, is fixed at the steady state. Furthermore, 
	Eq.~\ref{eq6} recover the steady state active power balance equation, $\mathbf{P} = \mathbf{B} \bm{\mathcal{A}}^\dagger \mathbf{B}^T \boldsymbol{\Delta}^*$ where $\boldsymbol{\Delta}^*$ is the GEDF at the steady state since $\mathbf{p}^* = \mathbf{P}$, $\mathbf{P}$ denotes the active powers of all nodes and $\bm{\psi}^* = \sin (\mathbf{B}^T\boldsymbol{\theta}^*) = \mathbf{B}^T \boldsymbol{\Delta}^*$.
	When $\mathbf{p}^*=\mathbf{P}$ at the steady state, the injected power at the generator node is equal to the load node. The vector $\boldsymbol{\Delta}^*$ reflects the steady-state power allocation at each node.
	
	\section{Contrastive Learning with Graph Embedding Dynamic Features}
	\label{GEDFL}
	
	% 框架图
	\begin{figure*}[tb]
		\centering
		\includegraphics[width=\textwidth]{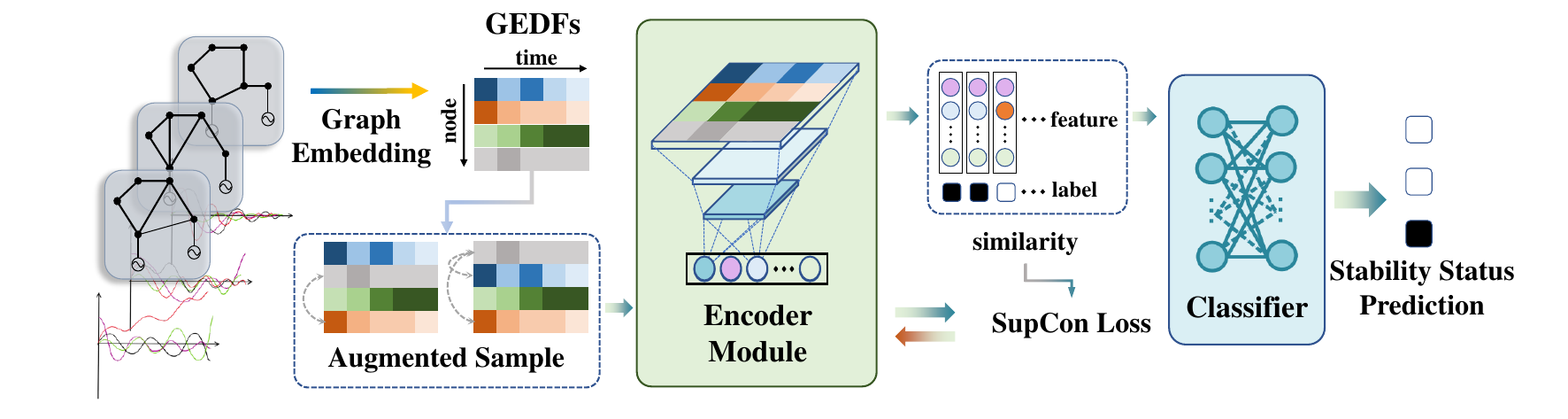}
		\caption{The framework of the transient stability GEDF-based supervised contrastive learning (GEDF-SCL).}
		\label{framework}
	\end{figure*}
 
	While data from different power systems exhibits heterogeneous distributions, more universal features should exist for TSA. Supervised contrastive learning~\cite{khosla2021supervised} can mitigate the influence of heterogeneous data distributions, bring data of the same label closer in feature space, and learn more universal features for transient stability assessment, enabling consistent predictions across diverse power system data. Supervised contrastive learning extends the self-supervised batch contrastive method ({\it{e.g.}} \cite{10.5555/3524938.3525087}) to a fully supervised setting, effectively capitalizing on label information. Random augmentations ("views") are created for the original data and used as input. Training is performed so that clusters of points belonging to the same class are pulled together in the embedding space while clusters of samples from different classes are pushed apart. For a random sample $i\in I\equiv{1,\cdots, M}$, let $j(i)$ be the augmented sample from the same original sample. Let $R(i) \equiv I\backslash {i}$ be the set excluding $i$ from all data. The loss function for supervised contrastive learning is defined as follows:
	\begin{equation}
		\label{SupCon_loss}
		\mathcal{L}^{sup} = -\sum_{i\in I} \frac{1}{|Pos(i)|} \sum_{m\in Pos(i)} \log \frac{e^{z_i\cdot z_{m}/\tau}}{\sum_{a\in R(i)}e^{z_i\cdot z_a/\tau}},
	\end{equation}
	where $z_l$ is the output of the encoder, $Pos(i)\equiv \{p\in R(i)|\tilde{y}_p=\tilde{y}_i \}$, which stands for all samples that share the same label with $i$, and $|P (i)|$ is its cardinality.
	
	Data augmentation is done by randomly changing the IDs of the nodes, {\it{i.e.}}, for $(n \times N)$ GEDFs, all vectors are randomly disrupted according to the first dimension (representing different nodes, from 1 to n). This method is designed to avoid making the model learn the wrong information if the fault always occurs at the same location. For example, if bus $k$, where $k\in \mathcal{V}$, consistently falls out of synchronization, the model may disproportionately focus on the data in the $k$-th row throughout the training process. Consequently, the model's generalization ability could be compromised.
	
	GEDF-SCL model has been developed using GEDFs as input to achieve accurate and highly adaptive transient stability prediction in power systems with alternating network topologies, as shown in Fig.~\ref{framework}. The model consists of three components: GEDFs construction and data augmentation, encoder module, and classifier.
	
	First, extensive transient operational data is obtained through time domain simulation. A feature fusion algorithm is then applied to incorporate topological structure information into this data, generating the GEDFs. Data augmentation is subsequently performed on the GEDFs using the aforementioned method.
	Second, the encoder is trained using supervised contrastive learning, encoding the input GEDFs into feature vectors of specific dimensions ($D_E=64$), which are then used by the classifier for prediction. Since GEDFs are not subject to spatial constraints based on topology, a two-dimensional CNN is applied directly for feature extraction. The network consists of three $3\times 3$ convolutional layers and a max pooling layer. The input first passes through two convolutional layers, followed by a pooling operation, and then another convolutional layer. Finally, a fully connected layer is used to produce an output of a given dimension ($D_E$).
	Lastly, the encoder outputs are fed into a fully connected multilayer perceptron (MLP) classifier for classification. The input is the output of the encoder module with dimension $D_E$, and the two intermediate layers have dimensions of 512 and 128, respectively. A GELU activation\cite{hendrycks2023gaussian} function is used in each layer.

	\section{Transient Stability Prediction and Alternating Grid Topology}
	\label{N-1_test}
	In order to prove the efficiency of the suggested approach, the IEEE 39-bus system was used as a basis for creating data and evaluating the ability to predict transient stability. The simulation of the power system's transient operation was explained, along with the process used to build the different topology power systems. Rotor angle stability (RAS) is a dynamic occurrence that is generally associated with alterations in active power flows, which lead to angular separation between synchronous units within the system.
	
	\subsection{Topology Change Dataset}
\begin{figure}[htbp]
		\centering
		\includegraphics[width=\linewidth]{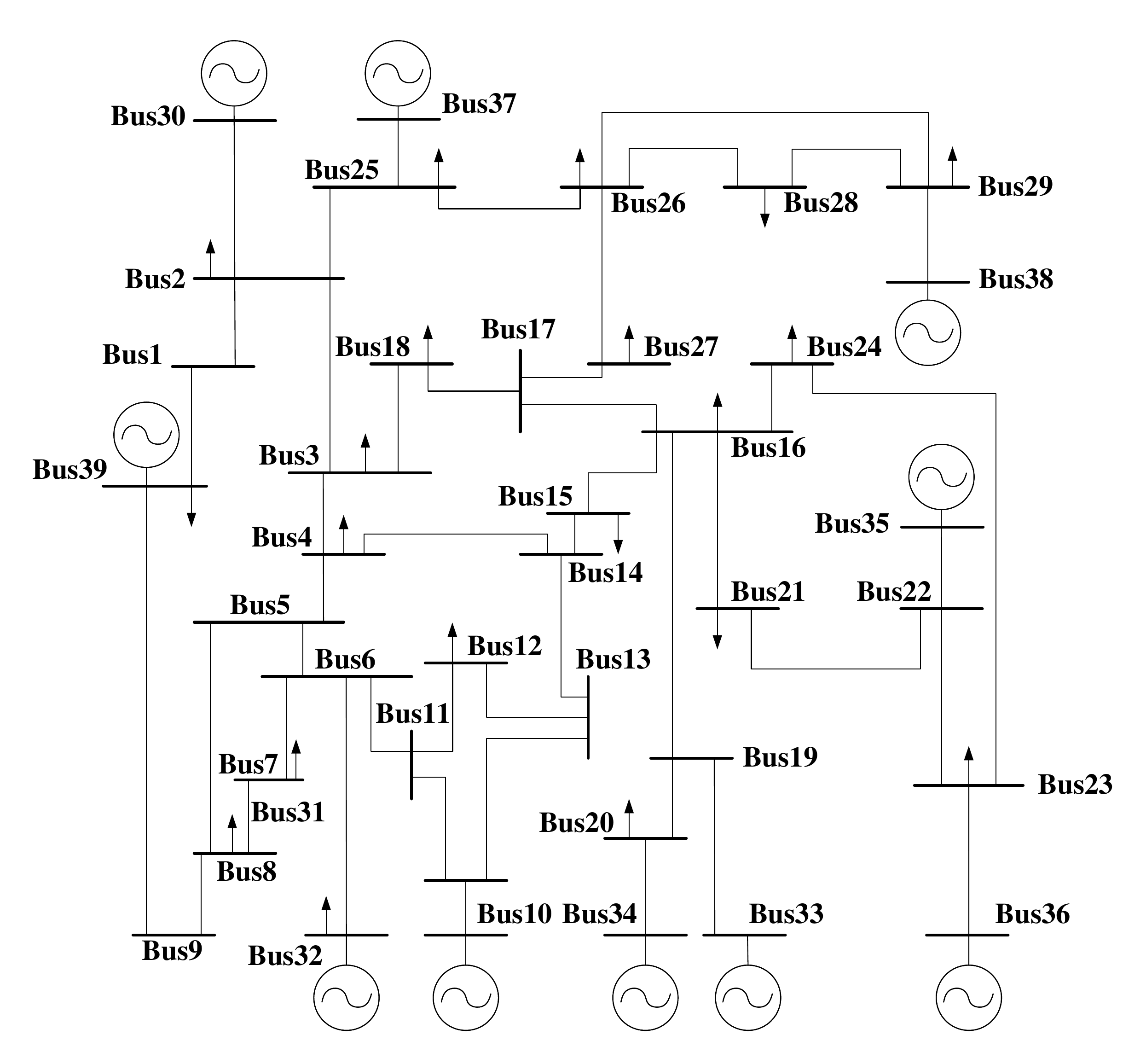}
		\caption{IEEE 39-bus system.}
		\label{IEEE-39}
	\end{figure}

	\paragraph{Topology Data} 
	The power system structure and device electrical parameters of the standard IEEE 39-bus system are used as the essential simulation environment. This includes 10 generators, 39 busbars, and 12 transformers. It is worth noting that all motors in the system are equipped with exciters, speed governors, and power system stabilizers (PSS) to ensure the stable operation of the power system. The topology is shown in Fig.~\ref{IEEE-39}. 
	
\begin{figure}[htbp]
	\centering
	\subfloat[New England IEEE 39-Bus System]{
		\label{fig2:subfig:a}
		\includegraphics[width=0.48\linewidth]{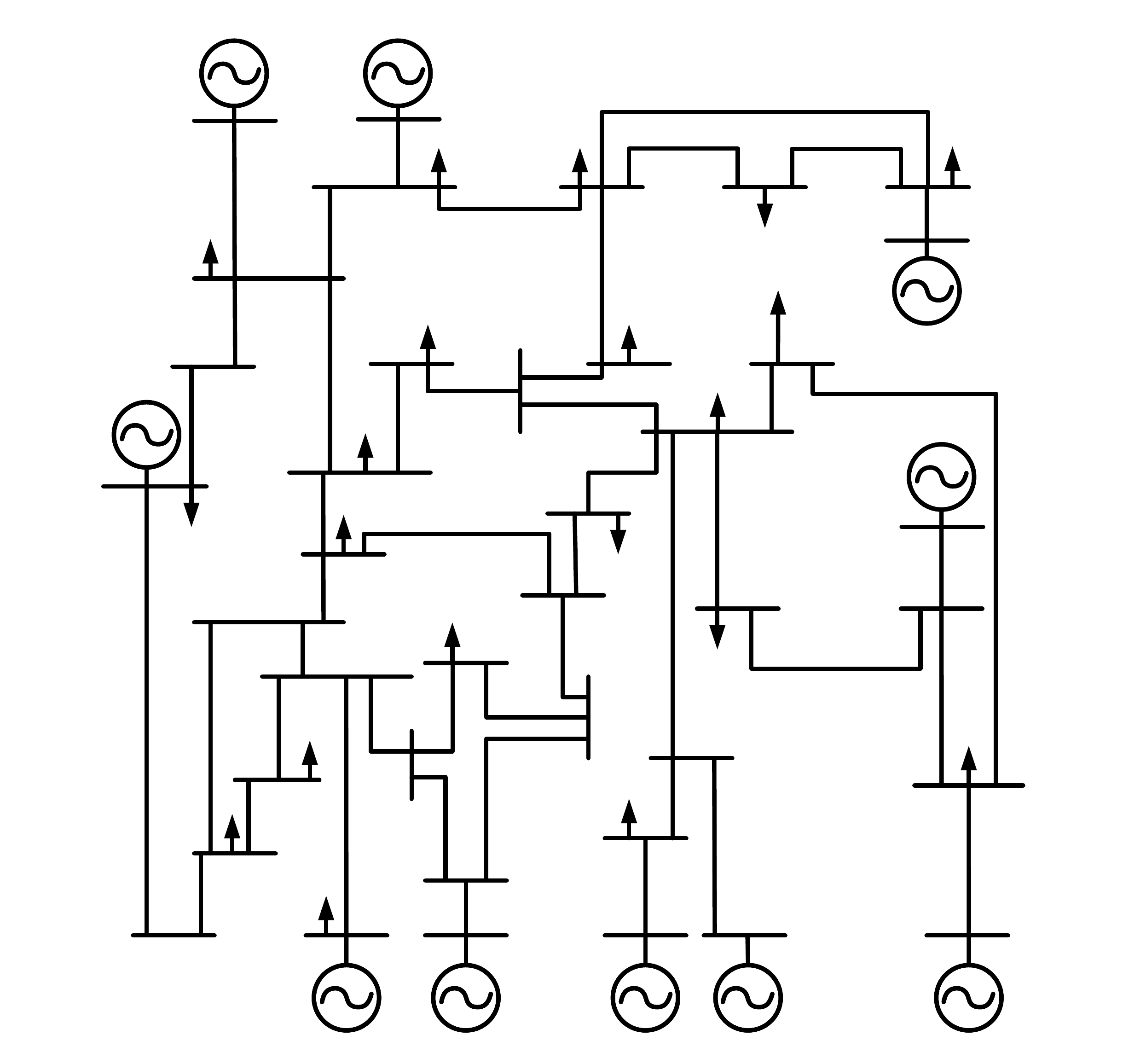}}
	\vspace{-0.1cm}
	\subfloat[Altered Topology 1]{
		\label{fig2:subfig:b}
		\includegraphics[width=0.48\linewidth]{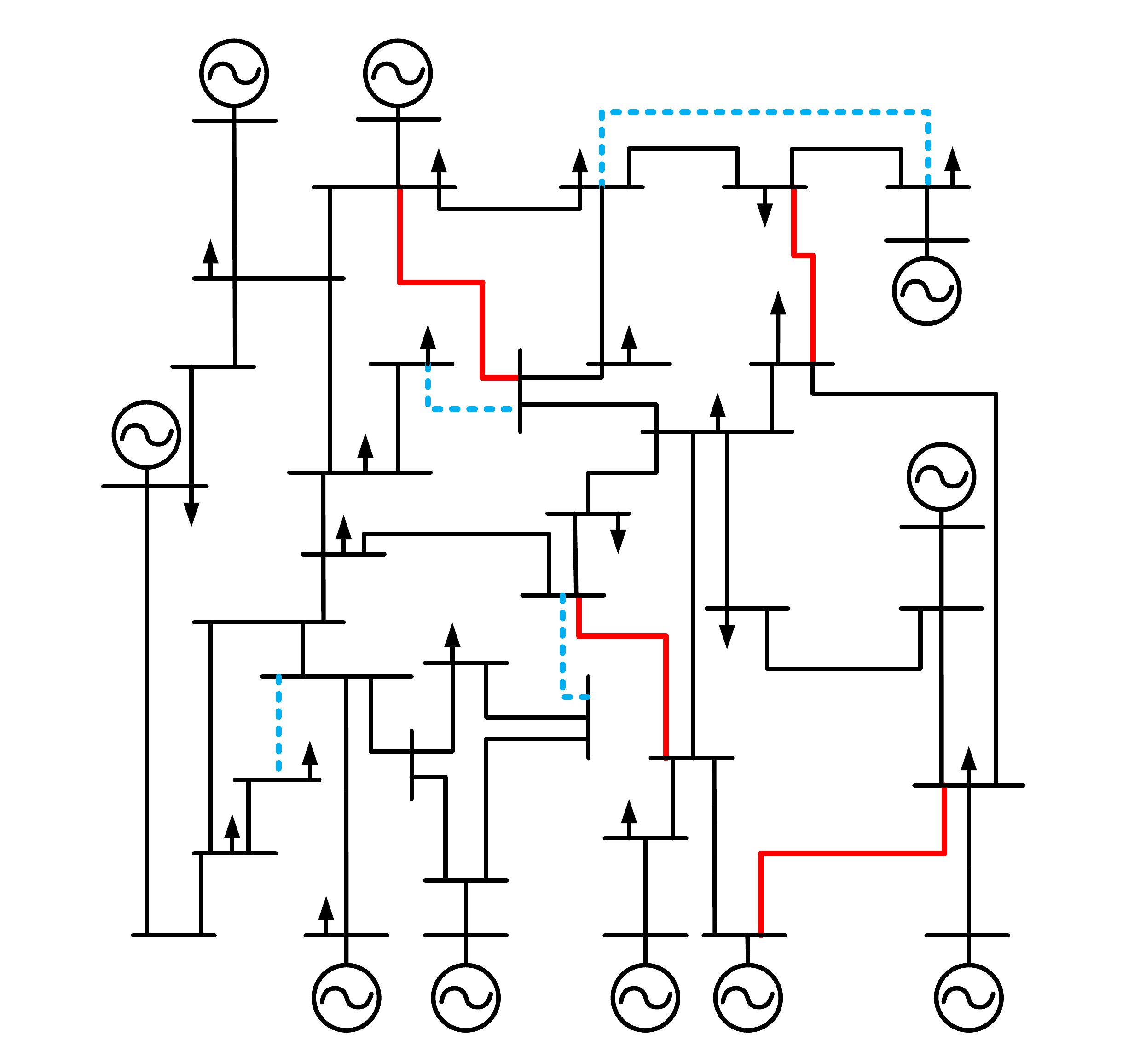}}
	\vspace{-0.1cm} 
	\subfloat[Altered Topology 2]{
		\label{fig2:subfig:c}
		\includegraphics[width=0.48\linewidth]{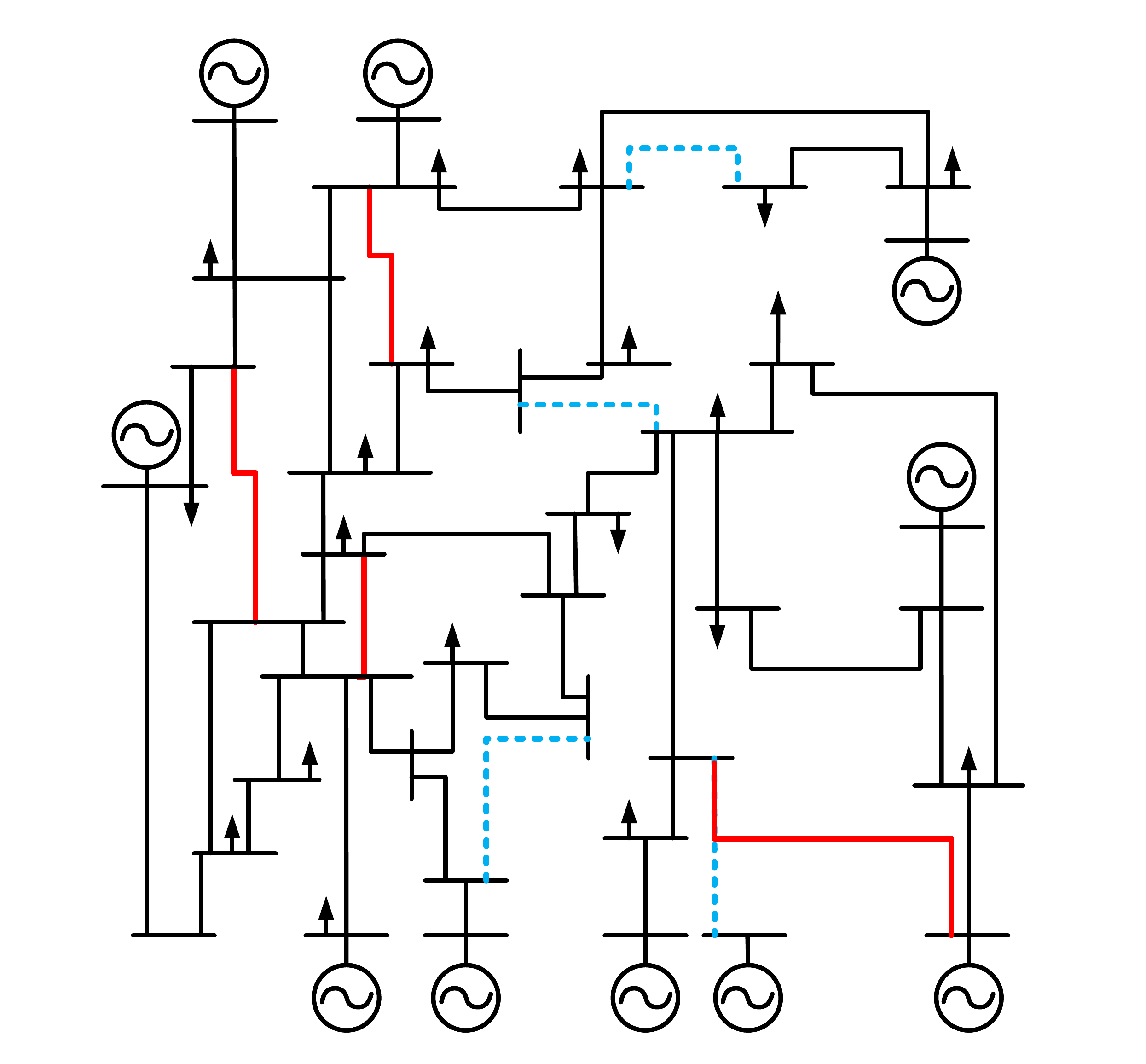}}
	\vspace{-0.1cm}
	\subfloat[Altered Topology 3]{
		\label{fig2:subfig:d}
		\includegraphics[width=0.48\linewidth]{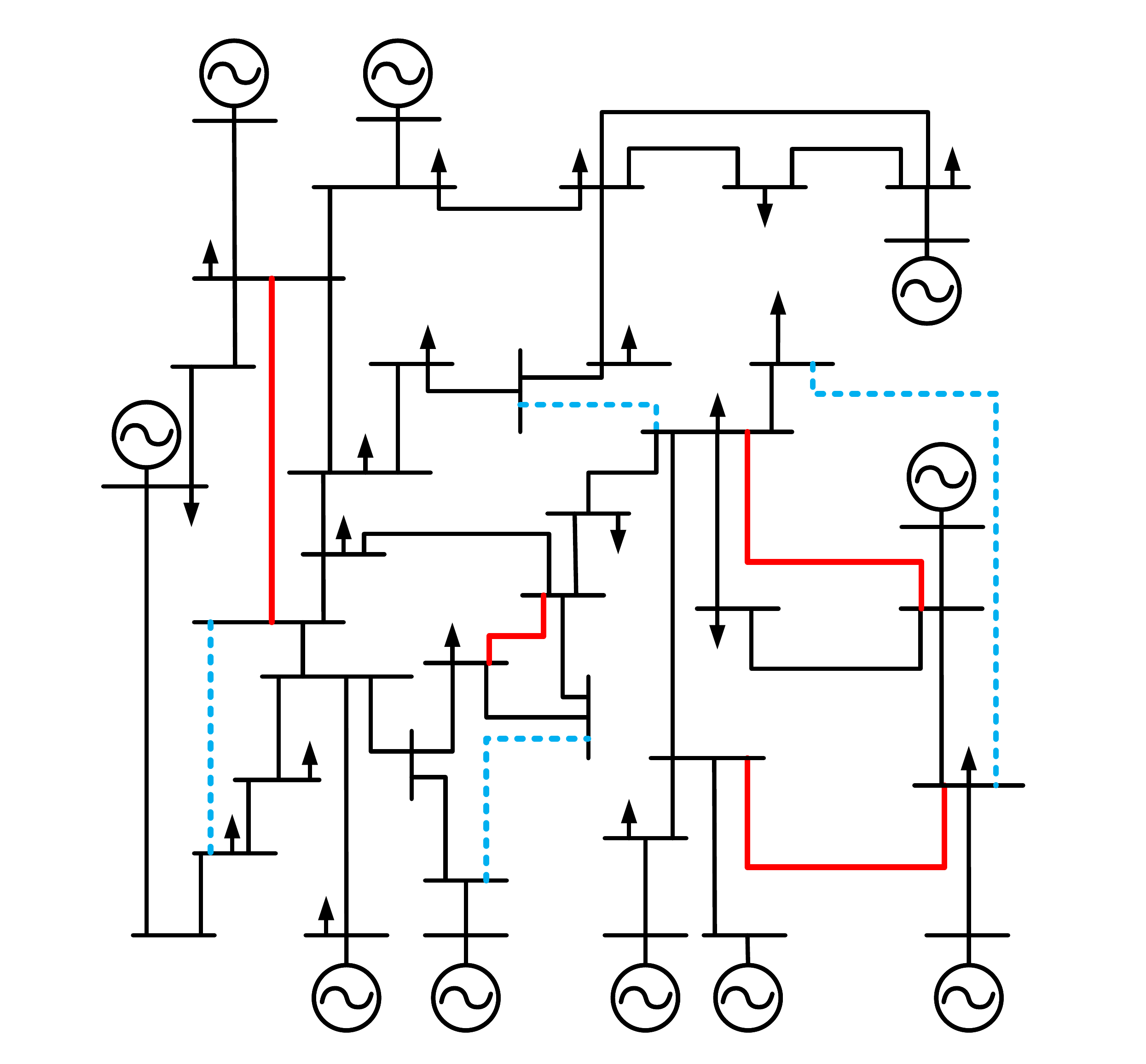}}
	\caption{\textbf{Illustration of grid toplogy changes.} Each new topology is generated by randomly removing four edges (shown as blue dashed lines in Fig.~\ref{fig2:subfig:b}\ref{fig2:subfig:c}\ref{fig2:subfig:d}) and simultaneously adding four new edges (shown as red solid lines).}
	\label{diff_topo}
	\end{figure}

	To alter the system topology, four edges were randomly removed from the IEEE 39-bus system, followed by adding edges elsewhere to connect four pairs of previously unconnected nodes. This provides a simple topology alteration without changing the total number of nodes and edges. Fig.~\ref{diff_topo} shows example topology alterations. Since the alterations involve random edge swaps, issues like isolated nodes or mismatched electrical parameters can occur. Therefore, when performing the alterations, it is necessary to check: (I) power system connectivity, requiring the new topology to be fully connected to avoid isolated nodes; and (II) whether the new topology can achieve a reasonable power flow solution, satisfying nodal power demand without exceeding equipment limits. 
	
	\paragraph{PMU Data} 
	Power outages and cascading failures (CF) have long been critical issues in power system operations. The N-1 criterion stipulates that when any component ({\it{e.g.}}, line, generator, transformer) disconnects under normal conditions, the system should maintain stability and power supply. Furthermore, other components should not be overloaded, and the voltage and frequency should be within an acceptable range\cite{phdthesis}. Accurately predicting system synchronization within a short time after faults would hold major significance for grid maintenance and outage response.
	
	In addition to the power grid topology change, nodal load disturbances were also incorporated by varying active and reactive loads in the generated PMU data. Specifically, the original load levels were randomly varied between 80\% and 120\% of base values, yielding 100 distinct scenarios.
	
	The power system toolbox (PST)\cite{PST} in Matlab was utilized to simulate power system transient dynamics. See the reference book\cite{PSTbook} for modelling details. The fault type, fault clearing time, simulation time, simulation duration, system base MVA, frequency and sampling rate were also set to mimic various historical conditions and contingencies. The detailed settings are listed in Table~\ref{init parameter}.
	
	\begin{table}[htbp]
		\centering
		\caption{CONDITION AND CONTINGENCY SETTING}
		\resizebox{\columnwidth}{!}{%
			\begin{tabular}{cc}
				\toprule[1.5pt]
				\textbf{Category}                    & \textbf{Shifts and Speciﬁcations}     \\ \midrule
				Fault types                          & 3-phase short circuit(N-1)            \\
				Fault application time               & $0.10s$                               \\
				\multirow{2}{*}{Fault clearing time} & $t_1=0.19s$ for near ends (random)    \\
				& $t_2=0.20s$ for remote ends           \\
				Simulation time                      & $10s$                                 \\
				\multirow{2}{*}{Timestep}            & $T_{step}=0.005s\quad(0\leq t\leq 2)$ \\
				& $T_{step}=0.01s\quad(2\leq t\leq 10)$ \\
				BaseMVA                              & 100                                   \\
				Frequency                            & 50Hz                                  \\
				\toprule[1.5pt]
			\end{tabular}%
		}
		\label{init parameter}
	\end{table}
	
\begin{figure}[h]
		\centering
		\includegraphics[width=\linewidth]{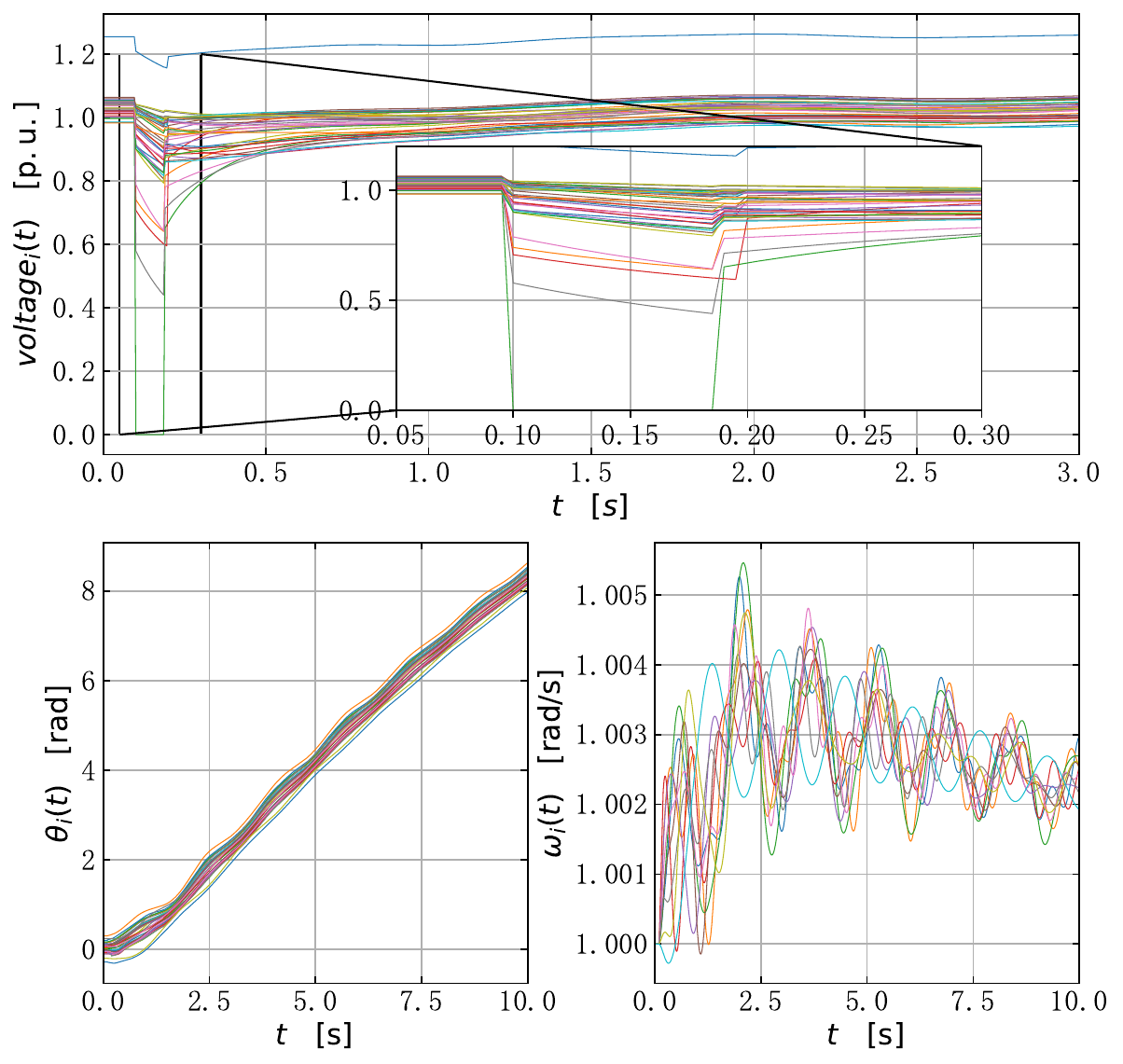}
		\caption{IEEE 39-bus N-1 simulations.}
		\label{n-1_dynamic}
	\end{figure}

\begin{figure}[htbp]
		\centering
		\subfloat[Stable Sample]{
			\label{Stable}
			\includegraphics[width=1\linewidth]{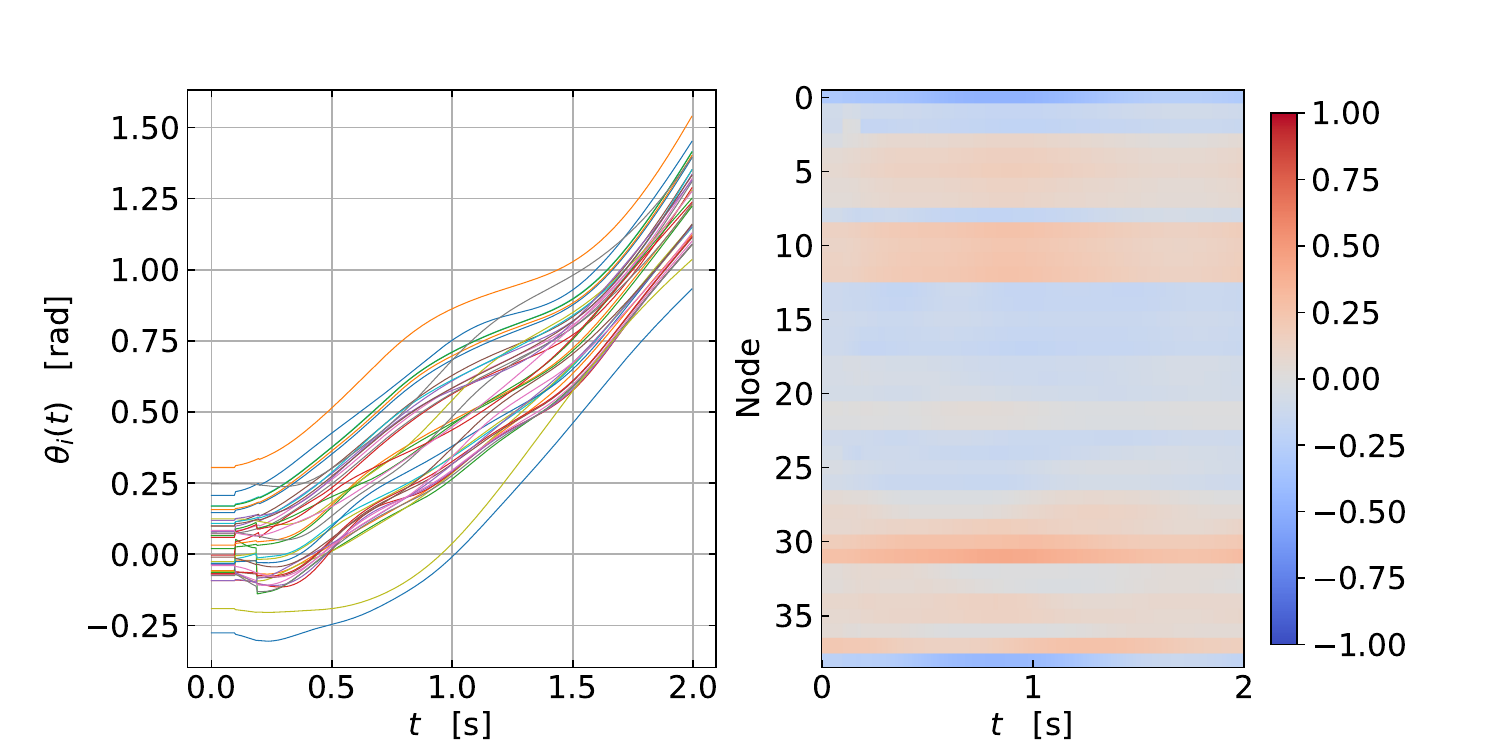}}
		\vspace{-0.1cm}
		\subfloat[Unstable Sample]{
			\label{Unstable}
			\includegraphics[width=1\linewidth]{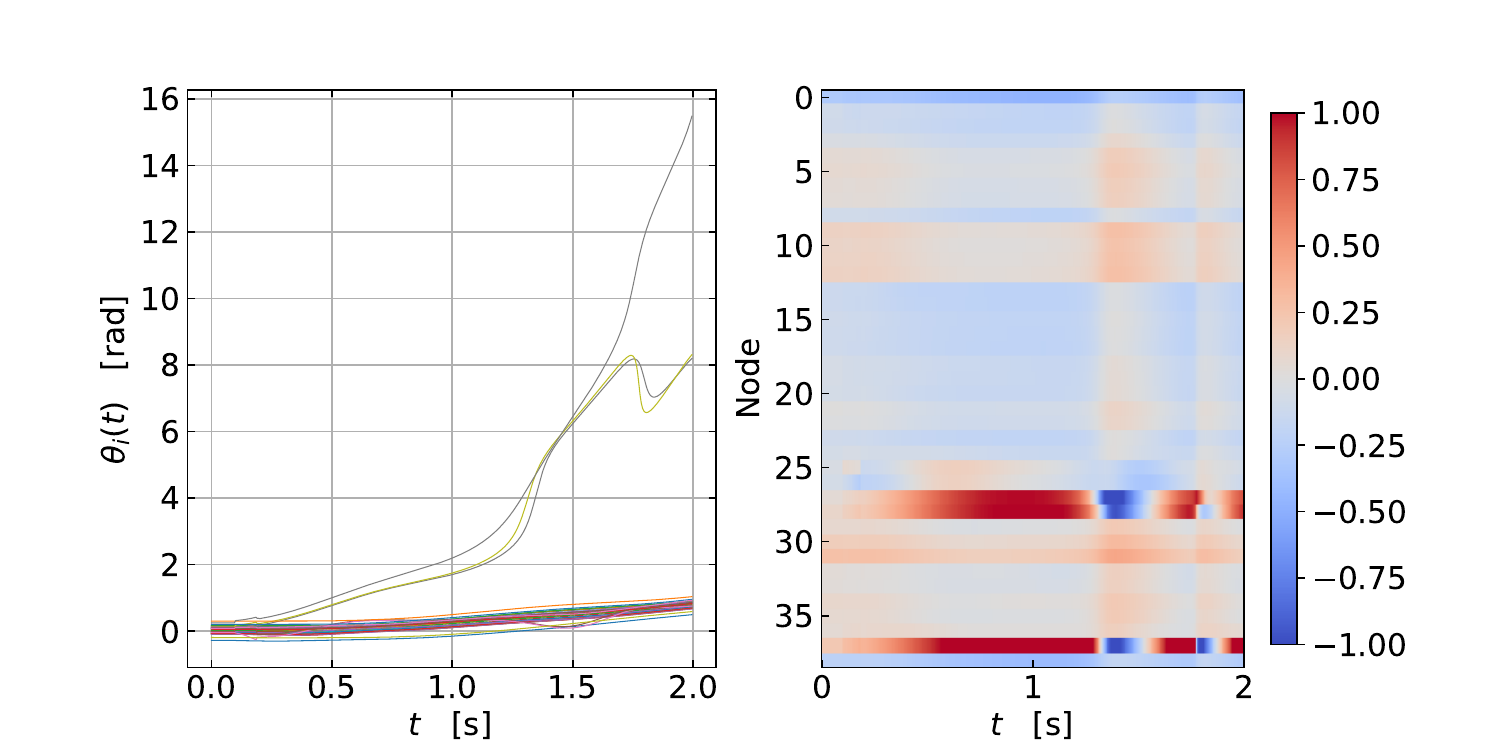}}
		\caption{Simulation results and its corresponding GEDFs.}
		\label{simulation result}
	\end{figure}

	The dynamic response obtained from the simulation is shown in Fig.~\ref{n-1_dynamic}. The GEDFs obtained under transient stability and instability are shown in Fig.~\ref{simulation result}. The difference between the two situations is clearly visible.
	
	In order to evaluate performance under different topological structures and fault conditions, it is essential to generate a significant number of scenarios for simulation purposes. In total, 317,679 datasets were generated with an approximate 2:1 synchronous to asynchronous ratio. Notably, the extensive training data was readily obtained through transient simulations, and the offline-trained model's runtime is unaffected by larger training sets. Each dataset consists of 0.05 s of post-fault phase angle data starting at 0.2 s fault clearing time. Approximately 70\% of the data (235,709 sets) were utilized for training, 10\% (26,190 sets) for validation. Two test sets containing about two thousand datasets were further partitioned. Test set 1 (T1) comprises new transient data from the topologies in the training set, while Test set 2 (T2) contains data from a completely new topology. That is, T1 represents new transient scenarios on the power systems which were encountered during training, while T2 represents entirely new power systems, further testing the model's generalization capability. Cross-validation was employed to evaluate prediction accuracy and avoid data dependency issues\cite{cross-validation}.
	
	\begin{table*}[bp]
		\centering
		\caption{Performance Comparison of Supervised Learning and Supervised Contrastive Learning on Test Datasets}
		\resizebox{2\columnwidth}{!}{%
			\begin{tabular}{cccccccccccc}
				\toprule[1.5pt]
				\multirow{2}{*}{\textbf{\begin{tabular}[c]{@{}c@{}}Training\\ Method\end{tabular}}} &
				\multirow{2}{*}{\textbf{Dataset}} &
				\multicolumn{5}{c}{\textbf{GEDFs}} &
				\multicolumn{5}{c}{\textbf{Raw Data}} \\
				&    & \textbf{ACC(\%)} & \textbf{Precision(\%)} & \textbf{Recall(\%)} & \textbf{F1-score(\%)} & \textbf{AUC} & \textbf{ACC(\%)} & \textbf{Precision(\%)} & \textbf{Recall(\%)} & \textbf{F1-score(\%)} & \textbf{AUC} \\ \midrule
				\multirow{2}{*}{\begin{tabular}[c]{@{}c@{}}Supervised\\ learning\end{tabular}}  & T1 & 93.26 & 93.81 & 96.77 & 95.27 & 0.9694 & 91.73 & 91.60 & 97.11 & 94.28 & 0.9562\\
				& T2 & 92.53 & 92.22 & 97.31 & 94.68 & 0.9602 & 90.95 & 90.24 & 97.34 & 93.64 & 0.9565 \\ \cmidrule(l){2-12} 
				\multirow{2}{*}{\begin{tabular}[c]{@{}c@{}}Supervised\\ contrastive learning\end{tabular}} & 
				T1 & 94.39 & 94.38 & 97.83 & 96.07 & 0.9735 & 92.78 & 93.25 & 96.70 & 94.95 & 0.9648 \\
				& T2 & 93.35 & 92.64 & 98.06 & 95.26 & 0.9656 & 92.07 & 92.17 & 96.69 & 94.34 & 0.9601\\ 
				\toprule[1.5pt]
			\end{tabular}%
		}
		\label{n-1_table}
	\end{table*}
	
	\subsection{N-1 Contingency Rotor Angle Stability }
	
	\subsubsection{Labeling Methodology}
	Transient stability index (TSI) is used to indicate the maintenance or loss of power system stability\cite{7445227}:
	
	\begin{equation}
		\label{label}
		\eta = \frac { 2\pi - \mid\Delta\delta_{max}\mid } {2\pi + \mid\Delta\delta_{max}\mid },
	\end{equation}
	$\Delta\delta_{max}$ is the absolute value of the maximum power angle separation between any two generators at the end of the post-fault system simulation. When the transient stability index $\eta > 0$, the system is considered stable and labeled as class "1"; otherwise, the system is classified as transiently unstable with label "0".
	
	\subsubsection{Training Results}
	The results section of the experiment compared and evaluated the performance differences between GEDFs and raw data. In addition, the supervised contrastive learning approach, GEDF-SCL, is investigated in comparison with the supervised learning, GEDF-SL, with the cross-entropy loss solely in the encoder module of the mode.
	
	Five metrics, ACC (accuracy), Precision, Recall, F1-score, and AUC (area under curve), were used to assess the performance of the model. Specifically, $\text{ACC}=\frac{TP+TN}{TP+FN+TN+FP}$, $\text{Precision}=\frac{TP}{TP+FP}$, $\text{Recall}=\frac{TP}{TP+FN}$, $\text{F1-score}=(1+\beta^2)\frac{Precision\cdot Recall}{\beta^2\cdot Precision + Recall}$, where TP, TN, FP, and FN denote true positives, true negatives, false positives, and false negatives respectively. AUC refers to the area under the receiver operating characteristic (ROC) curve bounded by the coordinate axes. By applying supervised learning and supervised contrastive learning methods to the GEDFs and raw data, respectively, the results are shown in Table~\ref{n-1_table}. As can be seen, the GEDFs derived from the power grid topology and transient dynamics information represent the nodal decoupled transient characteristic of the power system. Furthermore, supervised contrastive learning improves model generalization compared to supervised learning alone.
	
	In order to assess the effect of the input data's size on the model's effectiveness, the time length N described in Eq.~\ref{GEDF_def} was gradually increased from the previous 0.05 second to 0.2 second to test the model's predictions. Table~\ref{n-1_time} lists the performance characteristics of the models trained with data of these different time lengths.
	
	\begin{table}[h]
		\centering
		\caption{Effect of GEDFs input time length}
		\resizebox{\columnwidth}{!}{%
			\begin{tabular}{@{}cccccccc@{}}
				\toprule[1.5pt]
				\multirow{2}{*}{\textbf{\begin{tabular}[c]{@{}c@{}}Time\\ Length\end{tabular}}} &
				\multirow{2}{*}{\textbf{Dataset}} &
				\multicolumn{3}{c}{\textbf{GEDF-SL}} &
				\multicolumn{3}{c}{\textbf{GEDF-SCL}} \\
				&    & \textbf{ACC(\%)} & \textbf{F1-score(\%)} & \textbf{AUC} & \textbf{ACC(\%)} & \textbf{F1-score(\%)} & \textbf{AUC} \\ \midrule
				\multirow{2}{*}{0.05s} & T1 & 93.26 & 95.27 & 0.9694 & 94.39 & 96.07 & 0.9735\\
				& T2 & 92.53 & 94.68 & 0.9602 & 93.35 & 95.26 & 0.9656 \\ \cmidrule(l){2-8} 
				\multirow{2}{*}{0.10s} & T1 & 94.42 & 96.07 & 0.9748 & 95.32 & 96.72 & 0.9776\\
				& T2 & 93.67 & 95.43 & 0.9654 & 94.04 & 95.71 & 0.9669 \\ \cmidrule(l){2-8} 
				\multirow{2}{*}{0.15s} & T1 & 95.13 & 96.57 & 0.9795 & 95.80 & 97.05 & 0.9812\\
				& T2 & 94.49 & 96.02 & 0.9691 & 94.82 & 96.26 & 0.9687 \\ \cmidrule(l){2-8} 
				\multirow{2}{*}{0.20s} & T1 & 95.45 & 96.80 & 0.9821 & 96.22 & 97.34 & 0.9854 \\
				& T2 & 94.80 & 96.25 & 0.9732 & 95.23 & 96.56 & 0.9759 \\ \toprule[1.5pt]
			\end{tabular}%
		}
		\label{n-1_time}
	\end{table}
	
	Additionally, the classification performance of the proposed GEDF-SCL and other algorithms was compared on the same dataset to validate their effectiveness for transient stability assessment, with results summarized in Table~\ref{diff_model}.
	
	\begin{table}[htbp]
		\centering
		\caption{Results of different models}
		\resizebox{\columnwidth}{!}{%
			\begin{tabular}{ccccccc}
				\toprule[1.5pt]
				\multirow{2}{*}{\textbf{Model}} & \multicolumn{3}{c}{\textbf{T1}}                 & \multicolumn{3}{c}{\textbf{T2}} \\ \rule{0pt}{8pt}
				& \textbf{ACC(\%)}  & \textbf{F1-score(\%)} & \textbf{AUC} 
				& \textbf{ACC(\%)}  & \textbf{F1-score(\%)} & \textbf{AUC} \\ \midrule
				GCN\cite{kipf2017semisupervised} & 72.27 & 83.49 & 0.5995 & 67.07 & 79.52 & 0.6079 \\
				GAT\cite{GAT}  & 79.40 & 87.33 & 0.6761 & 72.47 & 82.21 & 0.6845 \\
				CNN (raw data) & 91.73 & 94.28 & 0.9562 & 90.95 & 93.64 & 0.9565 \\
				Transformer\cite{vaswani2017attention}    & 92.50 & 94.72 & 0.9585 & 92.08 & 94.34 & 0.9557 \\
				GEDF-SCL & \textbf{94.39} & \textbf{96.07} & \textbf{0.9735} & \textbf{93.35} & \textbf{95.26} & \textbf{0.9656} \\
				\toprule[1.5pt]
			\end{tabular}%
		}
		\label{diff_model}
	\end{table}
	
	The comparative results demonstrate Graph Neural Networks' difficulty in capturing effective features under significant topology changes. In contrast, CNN and Transformer models, which directly aggregate data without topology influence, exhibit improved performance. However, GEDF-SCL utilizes GEDFs encoding topology and transient data as input, enabling more effective feature classification learning through supervised contrastive learning. This achieves higher effectiveness with simpler model architecture, significantly outperforming more complex methods for transient stability assessment.
	
	As shown in Fig.~\ref{simulation result}, all data is normalized from -1 to 1. This standardizes the magnitude of the data and makes it easier to analyze. Secondly, the power calculation in Eq.~\eqref{P} correlates with phase angle differences between nodes, irreversibly mapping the angles from node space into edge space with some information loss. The GEDF vector is constructed from the original nodal data according to Eq.~\eqref{feature_map}, encoding the data with the power grid topology information. Here, all GEDFs are decoupled for the individual node. When analyzing the synchronous stability of generator nodes,  the localized GEDF vector of the generator nodes have the global power grid topology information.
	
	\section{Transfer Learning for N-$m$-1 Contingency}
	\label{N-$m$-1_test}
	The experiments above were carried out to obtain data on the transient dynamics of the power system with different topologies by modifying the topologies and slightly adjusting the node load. However, the topology alterations were limited, maintaining the same total edges and average node degree. The next step was to make a broader range of changes to the power system to test the model's transferability. The chosen transfer scenario is the N-$m$-1 contingency, {\it{i.e.}} a sequence of events consisting of the initial loss of $m$ generators or transmission components (primary contingency), followed by system adjustments, followed by a further loss of a single generator or transmission component (secondary contingency). The specific description is as follows:
	\begin{enumerate}
		\item{\textbf{Primary Contingency:} While ensuring the existence of the tidal equation solutions. Remove $m$ ($m=1,2,3$) edges while randomly varying load from 50-150\% of base, yielding 121 scenarios.}
		\item{\textbf{System Adjustments:} After a set of corrective actions performed automatically by the control system or manually by the system operator, the system reaches a new steady state.}
		\item{\textbf{Secondary Contingency:} In addition to the new steady state, generally due to accidents, another line is lost. More details on the generated datasets are provided in Table~\ref{transfer data}. Increasing $m$ makes finding stable cases more difficult, hence the smaller D3 size compared to D1 and D2.}
	\end{enumerate}
	
	\begin{table}[htbp]
		\centering
		\caption{N-$m$ Topology Change}
		\begin{tabular}{m{0.2\linewidth}<{\centering} m{0.2\linewidth}<{\centering} m{0.2\linewidth}<{\centering} m{0.2\linewidth}<{\centering}}
			\toprule[1.5pt]
			\textbf{Dataset} & \textbf{Number of Removed Edges ($m$)} & \textbf{Number of Generated Topologies} & \textbf{Total Generated Data}\\ 
			\midrule
			D1 & 1 & 15 & 14442 \\
			D2 & 2 & 15 & 11610 \\
			D3 & 3 & 10 & 5648  \\
			\toprule[1.5pt]
		\end{tabular}
		\label{transfer data}
	\end{table}
	
	The parameters of the encoder module are transferred, and the classifier is re-trained. The fine-tuning results are presented in Table~\ref{transfer_learn_new}. Only 20\% of the data is used for fine-tuning. The fine-tuning of the pre-trained model can significantly improve the efficiency of knowledge transfer.
	
	\begin{table}[htbp]
		\caption{Results from Fine-tuning}
		\resizebox{\columnwidth}{!}{%
			\begin{tabular}{@{}cccccccc@{}}
				\toprule[1.5pt]
				\multirow{2}{*}{\textbf{\begin{tabular}[c]{@{}c@{}}Training\\ Method\end{tabular}}} &
				\multirow{2}{*}{\textbf{Dataset}} &
				\multicolumn{3}{c}{\textbf{GEDFs}} &
				\multicolumn{3}{c}{\textbf{Raw Data}} \\
				&    & \textbf{ACC(\%)} & \textbf{F1-score(\%)} & \textbf{AUC} & \textbf{ACC(\%)} & \textbf{F1-score(\%)} & \textbf{AUC} \\ \midrule
				\multirow{3}{*}{\begin{tabular}[c]{@{}c@{}}Supervised\\ learning\end{tabular}} &
				D1 & 97.07 & 98.49 & 0.8972 & 96.61 & 98.25 & 0.8309 \\
				& D2 & 95.66 & 97.73 & 0.8770 & 95.57 & 97.71 & 0.8192 \\
				& D3 & 93.72 & 96.69 & 0.8112 & 92.87 & 96.25 & 0.7174 \\ \cmidrule(l){2-8} 
				\multirow{3}{*}{\begin{tabular}[c]{@{}c@{}}Supervised\\ contrastive\\ learning\end{tabular}} &
				D1 & 97.51 & 98.71 & 0.9236 & 96.20 & 98.05 & 0.8687 \\
				& D2 & 96.59 & 98.23 & 0.9008 & 95.70 & 97.78 & 0.8707 \\
				& D3 & 93.75 & 96.71 & 0.8288 & 93.08 & 96.38 & 0.7770 \\
				\toprule[1.5pt]
			\end{tabular}%
		}
		\label{transfer_learn_new}
	\end{table}
	
	The degree of topology change increases from dataset D1 to D3. Correspondingly, the improvement in effectiveness with GEDFs becomes more pronounced, with nearly 10\% higher AUC on dataset D3. This demonstrates GEDFs' strong capability in encoding topology information, yielding more generalizable feature representations to handle topology changes effectively.
	
	\section{Conclusion}
	\label{conclusion}
	The strong dependence on data and the fact that the trained model no longer fits new samples when the environment changes are the main problems that data-driven deep-learning approaches have to face. On the other hand, the power system is very versatile in terms of topology and device parameters. Therefore, there needs to be a method with stronger adaptability on TSP under changing topologies. The physics-informed GEDFs encode the transient data and power grid topology information. Using GEDFs as inputs, the GEDF-SCL is trained with supervised contrastive learning to predict the synchronous stability of the power system under transient conditions. The model can maintain high prediction accuracy when power system topology changes.
	
	The results show that GEDFs are more general features that allow the model to maintain predictive power in the face of power system changes caused by topology variations and load disturbances. Supervised contrastive learning can also enhance the model's classification ability. Under N-1 contingency, the proposed GEDF-SCL significantly outperforms graph neural networks and other more sophisticated neural network models. Even when dealing with unseen power grid topologies during training, it can still achieve stable and impressive predictive performance. Furthermore, by tuning the classifier, the model trained under N-1 contingency retains better predictive accuracy when transferred to N-$m$-1 contingency. Future work can further start from the physical principles of power systems to design more effective feature extraction algorithms and combine them with machine learning methods to realize transient stability assessment and prediction of power systems.
	
	\bibliographystyle{IEEEtran}
	\bibliography{reference.bib}
	
\end{document}